\documentclass{aa}
\usepackage{natbib}
\usepackage{graphicx}
\usepackage{color}
\usepackage[switch]{lineno} 
\bibpunct{(}{)}{;}{a}{}{,} 
\newcommand{\corot}{{\textsc{CoRoT}}}
\newcommand{\Kepler}{\emph{Kepler}}
\newcommand{\cible}{{HD\,175272}}
\newcommand\dnu{\Delta\nu}

\newcommand\numax{\nu\ind{max}}
\newcommand\nuc{\nu\ind{c}}
\newcommand\fonction{EACF}
\newcommand\ampmax{\mathcal{A}\ind{max}}
\newcommand{\ind}[1]{_{\mathrm{#1}}}
\newcommand{\Teff}{T\ind{eff}}
\newcommand{\Teffsun}{T\ind{eff\odot}}
\newcommand{\Teffref}{T\ind{eff,ref}}
\newcommand{\numaxsun}{\nu\ind{max\odot}}
\newcommand{\numaxref}{\nu\ind{max,ref}}
\newcommand{\gref}{g\ind{ref}}
\newcommand{\diff}{\mathrm{d}}
\newcommand{\Eqt}{Eq.}
\def\muHz{\,$\mu$Hz}
\def\mHz{\,mHz}
\def\m2s2{\,m$^{2}$\,s$^{-2}$} 
\def\kms{\,km\,s$^{-1}$}       
\def\vsini{$v\sin i$}          
\def\logg{\log g}
\def\Teff{T\ind{eff}}
\begin{document}
\title{Differential asteroseismic study of seismic twins observed by CoRoT\thanks{The CoRoT space mission,
launched on 2006 December 27, was developed and is operated by the
CNES, with participation of the Science Programs of ESA, ESA's
RSSD, Austria, Belgium, Brazil, Germany and Spain.}}

\subtitle{Comparison of HD 175272 with HD 181420}
\titlerunning{Asteroseismic twins}

\author{N. Ozel\inst{1,2}\and
B. Mosser\inst{3}\and
M.A. Dupret\inst{1}\and
H. Bruntt\inst{4}\and
C. Barban\inst{3}\and
S. Deheuvels\inst{5,12}\and
R.A. Garc{\'\i}a\inst{6}\and
E. Michel\inst{3}\and
R. Samadi\inst{3}\and
F. Baudin\inst{7}\and
S. Mathur\inst{8}\and
C. R\'egulo\inst{9,11}\and
M. Auvergne\inst{3}\and
C. Catala\inst{3}\and
P. Morel\inst{10}\and
B. Pichon\inst{10}
}
\offprints{B. Mosser}

\institute{Institut d'Astrophysique et de G\'eophysique de l'Universit\'e de Li\`ege, All\'ee du 6 Ao\^ut 17, 4000 Li\`ege, Belgium
\and Department of Astronomy and Space Science, Erciyes University, 38039 Melikgazi, Kayseri, Turkey
\and LESIA, CNRS, Universit\'e Pierre et Marie Curie, Universit\'e Denis Diderot,
Observatoire de Paris, 92195 Meudon cedex, France
\and Department of Physics and Astronomy, Aarhus University, DK-8000 Aarhus C, Denmark
\and Universit\'e de Toulouse, UPS-OMP, IRAP, Toulouse, France
\and Laboratoire AIM, CEA/DSM, CNRS - Universit\'e Paris Diderot, IRFU/SAp, 91191 Gif-sur-Yvette Cedex, France
\and Institut d'Astrophysique Spatiale, UMR 8617, Universit\'e Paris Sud, 91405 Orsay Cedex, France
\and Space Science Institute, 4750 Walnut street Suite\# 205, Boulder, CO 80301, USA
\and Instituto de Astrofisica de Canarias, 38205, La Laguna, Tenerife, Spain
\and Universit\'e de Nice-Sophia Antipolis, CNRS UMR 7293, Observatoire de la C\^ote d'Azur,
Laboratoire J.L. Lagrange, BP 4229, 06304 Nice Cedex 04, France
\and Universidad de La Laguna, Dpto de Astrofisica, 38206, La Laguna, Tenerife, Spain
\and NRS, IRAP, 14, avenue Edouard Belin, F-31400 Toulouse, France
}

\abstract{The CoRoT short asteroseismic runs give us the
opportunity to observe a large variety of late-type stars through
their solar-like oscillations. We report the observation and
modeling of the F5V star HD\,175272.}
{Our aim is to define a method for extracting as much information
as possible from a noisy oscillation spectrum. }
{We followed a differential approach that consists of using a
well-known star as a reference to characterize another star. We
used classical tools such as the envelope autocorrelation function
to derive the global seismic parameters of the star. We compared
HD\,175272\ with HD\,181420 through a linear approach, because
they appear to be asteroseismic twins.}
{
The comparison with the reference star enables us to substantially
enhance the scientific output for HD\,175272. First, we determined
its global characteristics through a detailed seismic analysis of
HD\,181420. Second, with our differential approach, we measured
the difference of mass, radius and age between HD\,175272 and
HD\,181420.}
{We have developed a general method able to derive
asteroseismic constraints on a star even in case of low-quality data.
This method can be applied
to stars with interesting properties but low signal-to-noise ratio
oscillation spectrum, such as stars hosting an exoplanet or
members of a binary system.}

 \maketitle \voffset
= 1.5cm
\section{Introduction}

Asteroseismology allows us to investigate the interior of stars.
The most detailed analysis of a star is based on the determination
of the largest possible number of oscillation frequencies, which
requires a high-quality oscillation spectrum
\citep[e.g.,][]{2010A&A...515A..87D,2012ApJ...748L..10M}. When
only global seismic parameters are determined, the output is
poorer, but nevertheless allows us to gain information that is not
given by classical spectrometric observations
\citep[e.g.,][]{2009A&A...506...41G,2010A&A...518A..53M}. The
output is limited for a low-quality oscillation spectrum. However,
because many interesting stars, for instance those hosting an
exoplanet, may present such a low-quality oscillation spectrum
\citep[e.g.,][]{2010A&A...524A..47G}, it is necessary to find a
method that optimizes the seismic information even in unfavorable
cases.

The seismic program of the CoRoT mission
\citep{2008Sci...322..558M} provides short runs in between
five-month-long runs, which allow us to study a larger set of
variable stars. HD\,175272, a solar-like star suspected to show
measurable solar-like oscillations, was a secondary target of the
first short run centered on HD\,175726
\citep{2009A&A...506...33M}. It was observed for 27 days in
October 2007. Despite the dim magnitude of the star and the
limited duration of the observation, we show that we can benefit
from the scaling relations observed in asteroseismology
\citep{1995A&A...293...87K} to enhance the accuracy of the
asteroseismic output. The comparison with a close reference star
with a high signal-to-noise (SNR) oscillation spectrum
\citep[HD181420,][]{2009A&A...506...51B} makes it possible to
benefit from the higher-precision models that can be derived from
the higher-quality oscillation spectrum.

Studying solar-like stars, solar analogs, or solar twins has
proved to be fruitful for investigating the influence of small
differences compared with the well-known solar case, as for
example $\tau$ Ceti \citep{2009A&A...494..237T}, $\alpha$\,Cen\,B
\citep{2005ApJ...635.1281K}, or 16 Cyg A and B
\citep{2012ApJ...748L..10M}. Similarly, studying 1-$M_\odot$
evolutionary sequences is of great interest
\citep{2011ApJ...740L...2S}. With the large increase of stars
showing solar-oscillations, we can now exploit the concept of a
differential seismic analysis between stellar twins and extend it
to references other than the Sun. We present this for the typical
case where, due to different SNR properties, a poorly constrained
star can benefit from the observations and the modeling of a
reference star.

The method presented here involves a differential analysis between
two stars with similar seismic properties. It avoids the possible
uncertainties caused by the extrapolation of the solar case, where
the Sun is used as a far-away reference. It is aimed at
constraining differences in internal physical processes of very
well constrained stellar twins, both seismically and
spectroscopically, from main-sequence stars to red giants. This
method helps avoiding the high inaccuracy in the forward-modeling
approach of poorly constrained stars. Indeed, when the parameter
space is too small, all models in this subspace significantly
differ from the real star. In contrast, a differential study is
less subject to systematic errors. An accurate measurement of the
differences between the target and a reference star is thus
possible.

In Section~\ref{etoile}, we discuss the physical parameters of
HD\,175272\ and the prediction of the asteroseismic signal by
scaling the star to a close reference, HD\,181420. Observations
are presented in Section~\ref{observations}. The analysis of the
power spectrum is analyzed in Section~\ref{signature}, with the
identification of the large separation and of its variation with
frequency. In Section~\ref{differential}, we first describe the
physics of our models and perform a seismic modeling of
HD\,181420, the reference star. We finally explain our
differential asteroseismic method and apply it to the study of the
second star HD\,175272. In Section~\ref{discussion}, we address
the  problem of using the frequency $\numax$ of the maximum
oscillation amplitude in scaling relations, especially for stars
that are not close to the solar type. We also examine how seismic
references can be defined. Section~\ref{conclusion} is devoted to
conclusions.


\section{Stellar parameters\label{etoile}}

HD\,175272, or HIP 92794, is known as an F5 dwarf. Its V magnitude
has been derived from Str\"omgren photometry by
\cite{1994A&AS..106..257O}. From  medium-resolution stellar
spectra, \cite{2001A&A...369.1048P} have estimated the effective
temperature to be about 6500\,K and $\logg = 4.09$; these
estimates are presented as poor. \cite{2004A&A...418..989N}, who
performed the Geneva-Copenhagen survey of the solar neighborhood,
have inferred the stellar mass to be about $1.44\pm0.06\, M_\odot$
and the metallicity about $-0.06$\,dex. They also estimated the
age to be about $1.8\pm 0.2$\,Gyr. \cite{2003A&A...406..203P} have
measured \vsini${} \simeq 23$\kms. They also found that
HD\,175272\ does not show any evident trace of variability.

We have revised the spectroscopic observations with a
high-resolution spectrum of HD\,175272\ recorded with the
spectrometer ELODIE at OHP. The updated values of $\Teff$ and
$\log g$ are given in  Table \ref{prop-phys}, with large
uncertainties related to the SNR of the recorded spectrum. The
abundances for 16 elements are given in Table~\ref{tab:ab}. They
were analyzed according to the method presented by
\cite{2004A&A...425..683B}, considering a microturbulence of
$1.70\pm0.23$\,km\,s$^{-1}$. This provides a mean metallicity
[Me/H] = $0.077\pm0.111$. This metallicity is the mean abundance
of metals with at least ten lines: Si, Fe, and Ni. The uncertainty
on [Me/H] includes the contributions from the uncertainties on
$\Teff$, $\log g$, and microturbulence.

According to these parameters, the seismic scaling relations can
be used to infer the expected global seismic parameters
\citep[e.g.,][]{2008Sci...322..558M,2011A&A...530A.142B,2011ApJ...742L...3W}.
The mean large frequency separation scales as the square root of
the mean density and is expected to be in the range [70, 90\muHz].
The frequency $\numax$ of maximum oscillation signal scales as the
acoustic cutoff frequency and is expected to be in the range [1.4,
1.9\,mHz]. The maximum bolometric amplitude of the radial modes is
expected to be in the range [3, 6\,ppm]
\citep{2007A&A...463..297S,2011ApJ...743..143H}.

\begin{table}
\caption{Primary parameters of HD\,175272}\label{prop-phys}
\begin{tabular}{lll}
\hline
                       & HD\,175272          & HD\,181420      \\
\hline
type                   & F5              & F2             \\
$T\ind{eff}$ (K)       & 6675$\pm$120    & 6580$\pm$100   \\
$[$Fe/H$]$             & $+0.08\pm0.11$  & $-0.05\pm0.06$ \\
$m\ind{V}$             & 7.43            & 6.57           \\
$L/L_\odot$            & 6.3$\pm$1       & 4.28$\pm$0.28  \\
$\Pi$   (mas)          & 11.82$\pm$0.95  & 20.21$\pm$0.94 \\
\hline
$\log g$ (cm\,s$^{-2}$)& 4.28$\pm$0.12   & 4.09$\pm$0.15  \\
$v \sin i$ (\kms)      & 23       & 18       \\
\hline
\end{tabular}
\end{table}

\begin{figure}
\centering
\includegraphics[width=8cm]{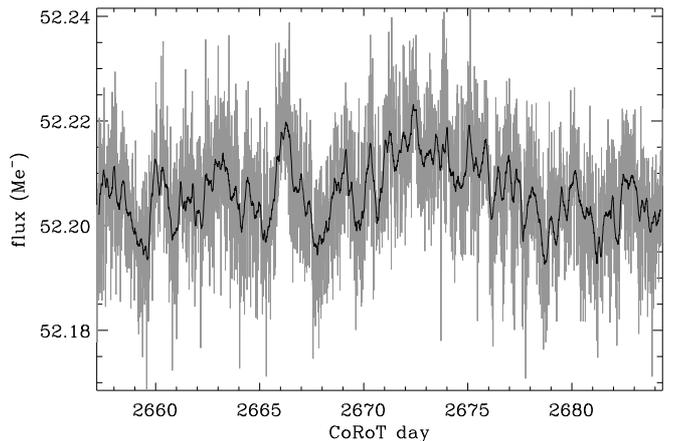}
\vskip 0.2cm \caption{Level-2 light curve of HD\,175272. The gray
curve shows a low-pass filtering, keeping one point per CoRoT
orbit. \label{intensity}}
\end{figure}

\begin{figure}
\centering
\includegraphics[width=8cm]{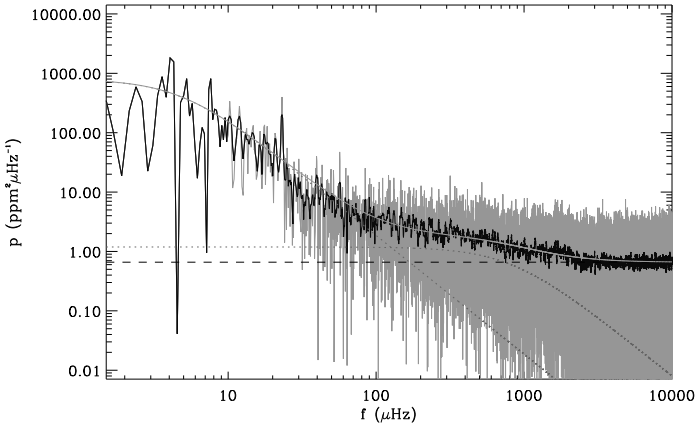}
\caption{Power density spectrum in log-scale axes. The peak around
4\muHz\ is probably the signature of the stellar rotation; the
peak at 23.2\muHz\ is an artifact due to the low-Earth orbit. The
black curve corresponds to a box-car-averaged spectrum, with a
varying smoothing window (the size of the window increases
linearly with frequency). The dashed line represents the
white-noise component; the background is modeled with two
Harvey-like components indicated by the dotted
lines.\label{logscale}}
\end{figure}

\begin{table}
 \centering
 \caption{Abundance of 16 elements in HD 175272.
 \label{tab:ab}}
 \begin{footnotesize}
\begin{tabular}{llrlr}
\hline
Element & Ab. & $N$ \\
\hline
  {C  \sc   i} &     $ -0.33\pm0.14$   &   1  \\
  {Na \sc   i} &     $ +0.22\pm0.14$   &   1  \\
  {Mg \sc   i} &     $ -0.05\pm0.14$   &   1  \\
  {Si \sc   i} &     $ +0.12\pm0.10$   &  10  \\
  {Si \sc  ii} &     $ +0.23\pm0.18$   &   2  \\
  {S  \sc   i} &     $ +0.06\pm0.14$   &   2  \\
  {Ca \sc   i} &     $ +0.17\pm0.11$   &   5  \\
  {Sc \sc  ii} &     $ +0.07\pm0.12$   &   2  \\
  {Ti \sc  ii} &     $ +0.11\pm0.12$   &   4  \\
  {Cr \sc   i} &     $ +0.09\pm0.12$   &   5  \\
  {Cr \sc  ii} &     $ +0.13\pm0.11$   &   4  \\
  {Mn \sc   i} &     $ -0.08\pm0.19$   &   4  \\
  {Fe \sc   i} &     $ +0.08\pm0.10$   &  91  \\
  {Fe \sc  ii} &     $ +0.14\pm0.10$   &  19  \\
  {Co \sc   i} &     $ +0.19\pm0.14$   &   1  \\
  {Ni \sc   i} &     $ +0.03\pm0.11$   &  13  \\
  {Cu \sc   i} &     $ -0.54\pm0.14$   &   1  \\
  {Zn \sc   i} &     $ -0.23\pm0.14$   &   1  \\
  {Y  \sc  ii} &     $ +0.06\pm0.14$   &   1  \\
  \hline
\end{tabular}
\end{footnotesize}
\end{table}
\section{Seismic observations\label{observations}}

\subsection{Time series}

This first short CoRoT run lasted 27.2 days in October 2007 with
HD\,175726 as the principal solar-like target
\citep{2009A&A...506...33M}. At the usual 32-s sampling of \corot\
seismic data, the whole time series includes 73\,426 points, and
the mean flux is about $5.22\cdot 10^7$ photoelectrons
(Fig.~\ref{intensity}).  The gaps due to data loss when the
satellite crossed the South Atlantic Anomaly are responsible for
the duty cycle of about 89.8\%, with 65\,944 original data points,
and the remainders obtained from interpolation.

When smoothed with one point per CoRoT orbit
(Fig.~\ref{intensity}), the time series shows rapid variation with
a typical amplitude of about 200\,ppm, much greater than the
expected standard deviation ($\simeq 10$\,ppm). The stellar origin
of these variations is most probable since similar features are
absent from the other times series recorded during the same CoRoT
run, even if the rapid variation with a period close to one day
occuring during the last third of the run may be instrumental
artifacts. The spot-modeling of the unperturbed light curve,
performed with the method developed by \cite{2009A&A...506..245M},
derives a surface rotation of the order of 2.8$\pm$0.4~days.
Compared with previously analyzed stars
\citep[e.g.,][]{2009A&A...506..245M,2011A&A...530A..97B}, we note
that the precision is limited by a poorer SNR.

\subsection{Low-frequency pattern}

The unfiltered power density spectrum of HD\,175272\ is
given in Fig.~\ref{logscale}. As presented in
\cite{2009A&A...506..411A}, it is affected by artifacts at the
orbital and diurnal frequencies. We had to correct these undesired
signatures, and performed a similar correction as in
\cite{2009A&A...506...33M}. Following \cite{2009A&A...495..979M},
we propose a fit for the stellar background component in the
low-frequency pattern, with two Lorentzian-like components in the
low-frequency range (below 1\mHz):
\begin{equation}
P(\nu) = \sum_{i=1}^2 {A_i\over 1 + \left(\displaystyle{\nu\over \nu_i}\right)^2}.
\label{lowfreqfit}
\end{equation}
Contrary to \cite{2009A&A...495..979M}, who introduced a
denominator varying as $\nu^4$, we note that an exponent of 2
provides a better fit. Because the time series shown in
Fig.~\ref{intensity} is quiet compared to HD\,175726, two
components are enough to provide an acceptable fit
(Table~\ref{low-freq}).

\begin{table}
\caption{Parameters of the two-component background.
}\label{low-freq}
\begin{tabular}{rll}
\hline
$\nu_i$ (\muHz)      & 4.84$\pm$0.30 &  798$\pm$47 \\
$A_i$ (ppm$^2$/\muHz)&  825$\pm$103  & 1.20$\pm$0.21 \\
\hline
\end{tabular}\end{table}

\subsection{Excess power at high frequency}

The high-frequency variations of the time series, after high-pass
filtering above the frequency range where oscillations are
expected, present a standard deviation of about 154\,ppm, in
agreement with the 132\,ppm value expected from pure photon noise
for such a star. Photon-noise-limited performance gives an
observed high-frequency power density of about
0.56\,ppm$^2$\muHz$^{-1}$, in agreement with the expected value.

\begin{figure}
\centering
\includegraphics[width=8.5cm]{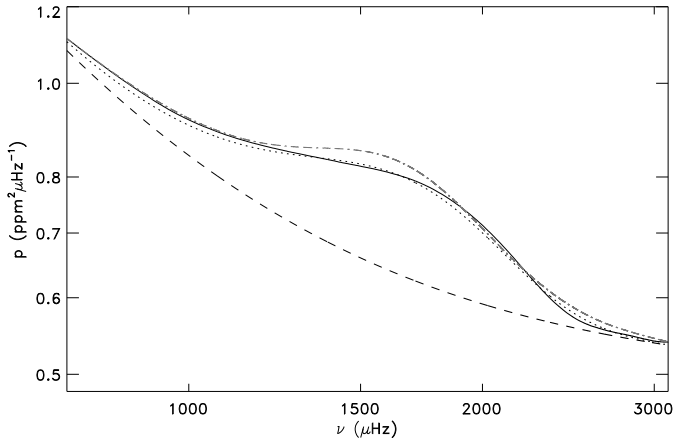}
\caption{Smoothed power density spectrum (with an apodized
300-$\mu$Hz FWHM window) in log-scale axes. The dashed line
represents the contributions of the granulation signal and of the
photon noise. The dotted line is a Gaussian fit of the excess
power envelope. The mixed line is the smoothed spectrum of
HD\,181420, with the same treatment and an offset accounting for
the different photon-noise levels. \label{smoothspectrum}}
\end{figure}

A strong smoothing of the spectrum with an apodized 300-$\mu$Hz
window was applied to show the evidence of excess power around
1.6\,mHz (Fig.~\ref{smoothspectrum}), as determined with several
methods \citep{2011MNRAS.415.3539V}. In the expected range, this
signature cannot be confused with the low-frequency contribution
described by Eq.~(\ref{lowfreqfit}). It corresponds to a
height-to-background ratio of about 45\,\%, which compares the
asteroseismic power with the background power at $\numax$, both
integrated over one large separation. We note that the excess
power in the spectrum of HD\,175272\ is similar to what has been
observed in HD\,181420 after correcting for the continuous
background levels caused by the photon noise. The locations of the
maximum power and its amplitude coincide.

\begin{figure}
\centering
\includegraphics[width=8.5cm]{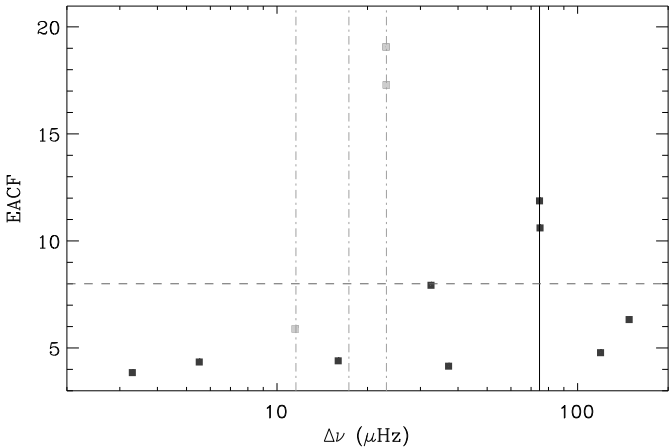}
\caption{Automatic determination of the large separation. Dark
squares indicate the values of \fonction\ tested in 13 frequency
ranges between 2 and 220\muHz. Gray values are artifacts cause by
the CoRoT low Earth orbit and must be excluded. The horizontal
dashed line indicates the threshold level for a detection at the
1\,\% rejection level. The vertical dark gray line indicates the
observed signature at $\dnu\simeq 75$\muHz, and the dash-dotted
lines the spurious signatures of the daily aliases (11.6 and
23.2\muHz). \label{fig_dnuauto}}
\end{figure}

\section{Data analysis\label{signature}}

\subsection{Mean seismic global parameters\label{mean}}

Despite the observed power excess, the Fourier spectrum does not
exhibit the regular pattern expected for solar-like oscillations
around 1.6\,mHz (Fig.~\ref{logscale}). Therefore, we used the
formalism (\fonction) and the automated procedure for a blind
detection of the large separation developed by
\cite{2009A&A...508..877M}. According to the global seismic
parameters of the spectrum and to the scaling of the EACF, we
expect a maximum of the envelope autocorrelation amplitude
$\ampmax$ of about 20, hence a fully reliable detection of the
large separation. We found a maximum amplitude $\ampmax\simeq 12$
(Fig.~\ref{fig_dnuauto}), lower than expected, but above the 1\%
rejection level, which is at eight for a blind detection of
solar-like oscillations in solar-like stars
\citep{2009A&A...508..877M}.

The blind analysis was followed by a more detailed study. The most
precise value of the mean large separation was derived from its
measurement in a broad frequency range around $\numax$, with a
filter of the same width as the envelope where oscillations are
detected. We measured 74.9$\pm$0.4\muHz. Other methods, similar to
those used by \cite{2011MNRAS.415.3539V}, converge on the same
value. Comparison of different methods has shown that the EACF
provides reliable results. The recent work by
\cite{2013A&A...550A.126M} and \cite{2013MNRAS.434.1668H} helps in
understanding this: the measurement of the large separation in
global conditions over a broad radial-order range is less
sensitive than local methods to the influence of the glitches
superimposed on the regular agency of the oscillation pattern.
Local methods can be precise, but do not necessary reach accuracy.

In similar conditions, the mean large separation of HD\,181420 is
75.2$\pm$0.04\muHz. The ratio between the two stars, defined as
the comparison of the low-SNR target with the reference, is about
0.996$\pm$0.005. Accordingly, we can derive that the stars have
very similar mean densities.

The maximum amplitude in the oscillation spectrum, estimated with
a Gaussian fit of the energy excess envelope, is at
1.60$\pm$0.03\,mHz. With the maximum for HD\,181420 reported at
1.61$\pm$0.01\,mHz, we derive a ratio of about 0.994$\pm$0.020.
This ratio is close to the ratio of the large separations. This
agrees with the scaling relation between $\numax$ and $\dnu$
reported in many stars
\citep[e.g.,][]{2009MNRAS.400L..80S,2010A&A...517A..22M}. We note
that these error bars are internal uncertainties only: they rely
on the assumption that the energy excess envelope has a Gaussian
form. Even if this is an usual assumption, it does not rely on a
firm theoretical basis. For a more evolved star such as Procyon,
this form is clearly not verified \citep{2010ApJ...713..935B}. The
comparison of the  different methods used for measuring $\numax$
has shown that there are small systematic bias. Here, only
internal uncertainties are relevant, because we perform a
differential analysis on twin stars with the same method.
\begin{figure}
\centering
\includegraphics[width=8.5cm]{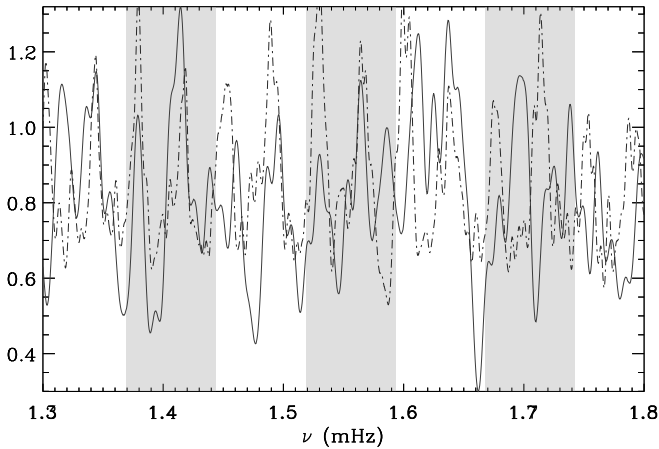}
\caption{Smoothed density spectrum of HD\,175272\ (Gaussian filter
of width 5\muHz). Gray and white domains indicate frequency ranges
with a width equal to the mean large separation. The dot-dashed
line gives the spectrum of HD\,181420, with frequencies multiplied
by a factor 0.996, also smoothed and corrected for the difference
between the white-noise contributions. \label{visimod}}
\end{figure}

\begin{figure}
\centering
\includegraphics[width=8.5cm]{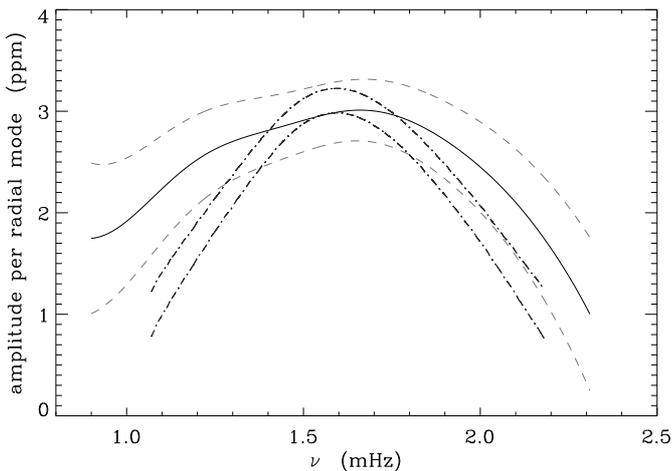}
\caption{Amplitude per radial mode determined according to
\cite{2009A&A...495..979M} for HD\,175272\ (solid line $+$ dashed
lines indicating the $\pm1-\sigma$ uncertainty) and for the
reference HD\,181420 (mixed lines indicating the uncertainty).
Error bars are principally due to the uncertainty in the power
density of the background. \label{power}}
\end{figure}

\subsection{P-mode pattern}

Owing to the low SNR, it is impossible to identify the oscillation
pattern precisely. A strong smoothing of the spectrum is required
to reveal it in the spectrum, which is incompatible with a precise
mode identification. The H0 test gives only a few eigenvalues at
the 10\,\% rejection level \citep{2004A&A...428.1039A}. Despite
this, the identification of the radial and dipole ridge is
possible, based on Figure~7 of \cite{2013A&A...550A.126M}. This
work shows that the observed $\varepsilon\ind{obs}$ term
describing the location of radial modes is mainly a function of
the ratio $\numax / \dnu$, modulated by the stellar mass, which is
also a function of the global parameters $\dnu$ and $\numax$. For
\cible, $\varepsilon\ind{obs}$ is about 1.0, so that the closest
radial mode to $\numax$ has a radial order $n=21$ and a frequency
$\nu_{21,0}  \simeq 1.57\,$mHz. The comparison with the Fourier
spectrum of HD\,181420, again based on a scaling factor of 0.996,
agrees with this identification (Fig.~\ref{visimod}).

\subsection{Mode amplitudes\label{amplit}}

The mode amplitude, determined according to the global recipe
reported in \cite{2009A&A...495..979M}, is about 3.0$\pm$0.3\,ppm.
The major contribution to the uncertainty comes from  the photon
noise and granulation signal. This value is similar to the value
3.6\,ppm expected from \cite{2007A&A...463..297S}. This shows that
the difficulty of observing the oscillation pattern of HD\,175272\
is mainly due to its faint magnitude (HD\,175272\ is twice as far
as HD\,181420) and to a limited observing run (one month versus
five months).

\begin{table}
\caption{Seismic constraints and their standard errors for HD\,181420 and HD\,175272. 
}\label{param175272}
\begin{tabular}{rll}
\hline
                    & HD\,175272       & HD\,181420 \\
\hline
$\dnu$  ($\mu$Hz)& 74.9$\pm$0.4 & 75.20$\pm$0.04   \\
$\numax $  (mHz)    & 1.60$\pm$0.03& 1.61$\pm$0.01 \\
\hline
\end{tabular}\end{table}

\section {Differential seismic modeling for seismic twins\label{differential}}

\subsection{Stellar models}\label{models}

Models were constructed using the CESAM2k stellar evolution code
\citep{morel97,morel08} for stellar structure and evolution. To
establish the oscillation frequencies, we used the adiabatic
oscillation code LOSC \citep{scuflaire08}. Following the
identification of solar-like oscillation properties of HD\,181420
\citep{2009A&A...506...51B}, modes with $n = 16-25$ and  $\ell
\leq 3$ were computed.

\begin{table*}
\begin{center}
\caption{Three different cases for modeling HD\,181420}
\label{IIIcases} \footnotesize{
\begin{tabular}{cccccccc}\hline
Case & Observational constraints &
\multicolumn{4}{c}{Model parameters} & \multicolumn{2}{c}{Outputs}  \\
& & $M / M_\odot$ & $t$ (\,Myr) & $Y_{0}$  & $\alpha$ & $R/R_{\odot}$& $L/L_{\odot}$ \\  \hline

I  & $\dnu$, $\numax$                        & 1.37 & 1942 &      &      & 1.63 & 4.58 \\
II & $\dnu$, $\numax$, $\Teff$               & 1.37 & 1998 & 0.27 &      & 1.64 & 4.52 \\
III& $\dnu$, $\numax$, $\Teff$, $L/L_{\odot}$& 1.38 & 1945 & 0.26 & 1.41 & 1.64 & 4.40 \\
\hline
\end{tabular}
}
\end{center}
\end{table*}

We used the EFF equation of state \citep{1973A&A....23..325E}, the
OPAL opacity tables \citep{1996ApJ...464..943I}, complemented at
low temperatures with the opacities of
\citet{1994ApJ...437..879A}. The nuclear reaction rates were
computed using the NACRE compilation \citep{1999NuPhA.656....3A}.
The atmosphere was derived assuming a gray Eddington atmosphere.
The adopted physical description for the convective zone is the
standard mixing-length theory (MLT \citealt{1958ZA.....46..108B}).
We computed two grids of models, one assuming the mixture of
\citet{gn93}  and the other the more recent mixture of
\citet{ags05}, respectively denoted GN93 and AGS05 hereafter.
Diffusion was not considered in the computation of the models.

We divided the problem into two parts. In a first step, we found a
reference model that best satisfied the set of observational
constraints for HD\,181420. The second step was to characterize
the lower-SNR target HD\,175272\ by comparing it with the
reference star HD\,181420 through a differential analysis.

\subsection{Seismic modeling of the reference star HD\,181420}\label{model-HD181420}

In our search for the best model of HD\,181420, we considered
different sets of constraints. This allowed us to determine the
importance of each of them in determining the stellar parameters
and to identify the problems that might be related to some of
them.

In case I, we only used the seismic constraints ($\dnu,\numax$)
and two free parameters: the age $t$ and the mass $M$. The other
parameters ($Y, \alpha$) were fixed to solar-calibrated values and
we adopted the metallicity given in Table~\ref{prop-phys}. For the
relation between $\numax$ and structure parameters, we adopted the
scaling relation
\begin{equation}
\label{numax}
\frac{\numax}{\numaxsun} = \frac{g}{g_{\odot}}\left(\frac{\Teff}{\Teffsun}\right)^{-1/2},
\end{equation}
with the unbiased calibration provided by
\cite{2013A&A...550A.126M}. To our surprise, we did not succeed in
finding an acceptable solution in this simple case, since the
inferred mass and radius differed too much from the values
expected from the photometric and spectrometric parameters. This
problem and its origin will become clear after the study of the
other cases.

In case II, we took the three constraints ($\dnu, \numax, \Teff$)
and three parameters ($M, t, Y_0$). We did not consider the
luminosity constraint because of its large uncertainty with a
relative error of about 9 \% (see Table~\ref{prop-phys}). The best
solution found in this case was $M=1.58\,M_{\odot}$,
$t=1.47$\,Gyr, $Y_{0}= 0.1982$. However, the agreement between
model and observations is still not satisfying in this case. To
clarify this, we simplified the problem even more by adopting the
scaling relation for $\dnu$
\begin{equation}
\frac{\dnu}{\dnu_{\odot}} = \sqrt{\frac{\rho}{\rho_{\odot}}},
\end{equation}
knowing from the modeling by \cite{2011ApJ...742L...3W} that a
calibration factor close to 1 is justified here. Adopting the
effective temperature given in Table~\ref{prop-phys} and the two
seismic constraints then leads to the solution: $M =
1.58\,M_{\odot}$, $R = 1.69\,R_{\odot}$, which are unrealistic
values in view of the effective temperature and standard stellar
evolution theory.

Finally, in case III, we performed a minimization with four
constraints ($\Delta \nu,\numax$, $T_{\rm eff}$, $L/L_{\odot}$)
and four parameters ($M, t, Y_0, \alpha$): $M=1.53\,M_{\odot}$,
$t=1.46$\,Gyr, $Y_{0}= 0.1948$, $\alpha = 1.05$. However, some of
the stellar parameters we found, such as the mass, the radius, and
the luminosity, were unrealistic for solar-like stars.

These results are summarized in Table~\ref{IIIcases}. As a
consequence of this preliminary study, a clear difficulty appeared
to be finding a model that reproduced a set of constraints
including $\numax$. This could come from an inaccurate measurement
of $\numax$  or could indicate that the scaling relation
(\Eqt~\ref{numax}) for $\numax$ calibrated on the Sun is too
inaccurate for an F2 star such as HD\,181420. Again, the case of
Procyon justifies that the scaling relations cannot be considered
to be accurate for such a star, since the seismic and modeled
masses differ by about 25\,\% \citep[see, e.g., Table 1
of][]{2013A&A...550A.126M}. We note that, because of the stellar
mass higher than 1.3\,$M_\odot$, the oscillation spectrum of
HD\,181420 is less accurately described by the asymptotic
expansion, the value of the offset $\varepsilon$ being
significantly lower than expected \citep{2013A&A...550A.126M}.
This may also explain why the scaling relations are not as
accurate as for lower-mass stars. The departure from both the
$\numax$ and $\varepsilon$ scaling relations is caused by the
significant changes of the physical properties of the external
layers with increasing effective temperature.

Hence, we finally chose to remove this constraint for the seismic
modeling of HD\,181420. However, we retained it for the
differential analysis (see next section). The final set of three
constraints adopted for determining the best model is thus $\dnu$,
$\Teff$, $L/L_{\odot}$ (see Tables \ref{prop-phys} and
\ref{param175272} for the observed values). We assumed as free
parameters the mass $M$, the age $t$, and the initial helium
abundance $Y$. For our models, we used the solar mixture of
\citet{gn93} as well as the new one of \citet{ags05}. We used a
solar-calibrated value of the mixing-length  parameter thoughout.
We determined the best models that minimized the $\chi^2$ fitting
function. Table~\ref{181420} gives their parameters for the two
chemical mixtures. To determine the uncertainty of the model
parameters of HD\,181420, we used the singular value decomposition
method (SVD; for a detailed description, see e.g.,
\citet{2011A&A...532A..82O} and references therein).

\begin{table}
\caption{Best models of HD\,181420. The upper part gives the model
parameters and the bottom part gives the theoretical  values of
the constraints found for these models.} \label{181420}
\begin{tabular}{cccc}
\hline
 & solar mixture GN93 & & solar mixture AGS05 \\
 \hline
$M_1$ & $1.30 \pm 0.17$ & & $1.28 \pm 0.17$ \\
$t_1$(\,Myr) & $2127 \pm 175$ & & $2325 \pm 267$ \\
$(Y_{0})_1$ & $0.30 \pm 0.09 $ & & $0.29\pm 0.09$ \\
$R/R_{\odot}$ &  $1.61 \pm $ 0.10 &  & $1.60 \pm 0.10$  \\
\hline
$\Delta \nu_{\rm th}$($\mu$Hz) & 75.2 & & 75.2  \\
$T_{\rm eff, th}$(K) & 6542 & & 6574 \\
$L/L_{\odot, \rm th}$ & 4.28 & & 4.29 \\
\hline
\end{tabular}
\end{table}

\subsection{Differential analysis for HD\,175272\ with the scaling relation for $\Delta \nu$}\label{Deltanu}

As indicated above, the second step was performing a differential
seismic analysis of the star HD\,175272, based on its similarity
with the reference star HD\,181420. From Table~\ref{prop-phys}, we
see that the estimated luminosities of the two stars are
different. This seems unrealistic in view of their very similar
other characteristics. Moreover, the luminosity error bar for
HD\,175272 is very large. We therefore decided to exclude the
luminosity difference as a constraint in our differential
analysis. The relative differences of observational constraints
used as input for our differential analysis are given in Table
\ref{diffobs}. These values are deduced from
Tables~\ref{prop-phys} and~\ref{param175272}. For $Z/X_{0}$,
slight differences are found depending on the adopted solar
mixture. With the subscript 1 referring to HD\,181420 and 2 to
HD\,175272, we have $(Z/X_{0})_1 = 0.0218 \pm 0.0030$ and
$(Z/X_{0})_2 = 0.0295 \pm 0.0075$ for the mixture of \cite{gn93}.
For the more recent mixture of \cite{ags05}, we have
$(Z/X)_{\odot} = 0.0165$, which leads to $(Z/X_{0})_1 = 0.0147 \pm
0.0020$ and $(Z/X_{0})_2 = 0.0198 \pm 0.0050$.

As previously mentioned, we decided to exclude the constraint
$\numax$ in obtaining the reference model of the star HD\,181420.
Also, both stars, HD\,181420 and HD\,175272, are very different
from the Sun in terms of their mass, radius, and luminosity.
Therefore, we assume that \Eqt~(\ref{numax}) is not accurate
enough for comparison of the two stars with the Sun. However,
these two stars are similar to each other with respect to their
seismic properties. In this situation we can expect that, although
the scaling relation is not valid for a comparison with the Sun,
it is valid for the comparison between two stars that have similar
properties,
\begin{equation}
\label{numax2}
\frac{\numax}{\numaxref} = \frac{g}{\gref}\left(\frac{\Teff}{\Teffref}\right)^{-1/2},
\end{equation}
where the index `ref' refers to the values of the reference star.
As a first step, we also assumed that the mean large separation
$\dnu$ is proportional to the square root of the stellar density,
$\dnu \propto \sqrt{M/R^3}$. We therefore used the following
scaling relations:
\begin{eqnarray}
\label{scaling1}
\frac{R}{R_{\rm ref}} = \left( \frac{\numax}{\numaxref}\right) \left(\frac{\dnu}{\Delta \nu_{\rm ref}}\right)^{-2} \left( \frac{\Teff}{\Teffref}\right)^{1/2}, \\
\label{scaling2}
\frac{M}{M_{\rm ref}} = \left(\frac{\numax}{\numaxref}\right)^{3} \left(\frac{\dnu}{\Delta \nu_{\rm ref}}\right)^{-4} \left( \frac{\Teff}{\Teffref}\right)^{3/2},
\end{eqnarray}
where $\Teff$, $R$, $M$ are the effective temperature, radius, and
mass of the stars, respectively. If $\dnu$, $\numax$ and $\Teff$
are known, Eqs.~(\ref{scaling1}) and (\ref{scaling2}) directly
yield the stellar mass and radius. From Eqs.~(\ref{scaling1}) and
(\ref{scaling2}), we obtain the differential equations
\begin{eqnarray}
\label{dRR}
\frac {\diff R}{R} = \frac{\diff \numax}{\numax} -2\frac{\diff \dnu}{\dnu} + \frac{1}{2}\frac{\diff \Teff}{\Teff}, \\
\label{dMM}
\frac{\diff M}{M} = 3\frac{\diff \numax}{\numax}- 4\frac{\diff \dnu}{\dnu} + \frac{3}{2}\frac{\diff \Teff}{\Teff},
\end{eqnarray}
where $\diff R/R$ and $\diff M/M$ are the relative difference in
radius and mass between the two stars, related to the relative
differences $\diff \numax/\numax$, $\diff \dnu/\dnu$ and $\diff
\Teff/\Teff$ given by the observations (Table \ref{diffobs}). The
solutions of Eqs.~(\ref{dRR}) and (\ref{dMM}) for these input
values are given in Table~\ref{diff1-results}. Note that the mass
and  radius differences found using scaling relations are
independent of stellar evolutionary models. Then, by
differentiating the relations $L=L(M,t,Y_0,Z/X_0)$ and $\Teff$ $=
\Teff(M,t,Y_0,Z/X_0)$ given by the stellar evolutionary tracks and
using Stefan-Boltzmann's law, $L\propto R^2\Teff^4$, we obtained
the following equations:
\begin{eqnarray}
\label{diffL}
\frac{\diff L}{L} &= &2\frac{\diff R}{R} + 4\frac{\diff \Teff}{\Teff}  =
\frac{\partial \ln L}{\partial \ln M}\frac{\diff M}{M} +
\frac{\partial \ln L}{\partial \ln t}\frac{\diff t}{t} \nonumber \\
&& + \frac{\partial \ln L}{\partial \ln Y_0}\frac{dY_0}{Y_0} + \frac{\partial \ln L}{\partial \ln Z/X_0}\frac{\diff Z/X_0}{Z/X_0}, \\
\label{diffTeff}
\frac{\diff \Teff}{\Teff}  &= & \frac{\partial \ln \Teff}{\partial \ln M}\frac{\diff M}{M} +
\frac{\partial \ln \Teff}{\partial \ln t}\frac{\diff t}{t}+ \frac{\partial \ln \Teff}{\partial \ln Y_0}\frac{dY_0}{Y_0} \nonumber \\
&& +\frac{\partial \ln \Teff}{\partial \ln Z/X_0}\frac{\diff
Z/X_0}{Z/X_0} .
\end{eqnarray}
The value of the first term $\diff R/R$ on the left-hand side of
\Eqt~(\ref{diffL}) and that of the term $\diff M/M$ are obtained
from previous step (Eqs.~(\ref{dRR}) and (\ref{dMM})), and the
term $\diff \Teff / \Teff$ is obtained from the observations.
Thus, we have a linear system of equations with two unknowns
($\diff Y_0/Y_0, \diff t/t$), which then allows us to determine
the differences in initial helium abundance and age between the
two stars.

\begin{table}
\caption{Observed relative differences between the two stars used
as input for the differential analysis.} \label{diffobs}
\begin{tabular}{lcl}
\hline
$\diff \dnu/\dnu \pm \sigma_{\dnu}$              &=& $-0.004 \pm 0.005$  \\
$\diff \Teff/\Teff \pm \sigma_{\Teff}$           &=& $0.014 \pm 0.023$ \\
$\diff \numax/\numax \pm \sigma_{\numax}$        &=& $-0.006 \pm 0.020$\\
$\diff (Z/X_{0})/(Z/X_{0}) \pm \sigma_{Z/X_{0}}$ &=& $0.300 \pm 0.282$ (GN93)\\
$\diff (Z/X_{0})/(Z/X_{0}) \pm \sigma_{Z/X_{0}}$ &=& $0.296 \pm 0.281$ (AGS05)\\
\hline
\end{tabular}
\end{table}

\begin{table}
\caption{Relative differences between the two stars deduced from our differential analysis, using the scaling
relation for $\dnu$.
Radius and mass relative differences are obtained from Eqs. (\ref{dRR}) and (\ref{dMM}). Age and initial helium abundance relative differences are obtained from Eqs. (\ref{diffL}) and (\ref{diffTeff}).} \label{diff1-results}
\begin{tabular}{lrr}
\hline
\multicolumn{3}{c}{$\diff R/R \pm \sigma_R = 0.009 \pm 0.025 $} \\
\multicolumn{3}{c}{$\diff M/M \pm \sigma_M = 0.019 \pm 0.072 $} \\
\hline
Solar mixture & GN93 & AGS05 \\
$\diff t/t \pm \sigma_{\diff t}$ & $ -0.27 \pm 0.29 $ & $  -0.28 \pm 0.31 $ \\
$\diff Y_{0}/Y_{0} \pm \sigma_{\diff Y_{0}}$ & $ 0.10 \pm 0.21$ & $ 0.13 \pm 0.24 $ \\
\hline
\end{tabular}
\end{table}

\begin{table}
\caption{Relative differences between the two stars obtained from full computations of adiabatic frequencies
and solving Eqs.~(\ref{diff3-1})-(\ref{diff3-3}).} \label{diff2-results}
\begin{tabular}{lrr}\hline
 solar mixture & GN93 & AGS05 \\
\hline
$\diff M/M \pm \sigma_{M}$ & $0.02 \pm 0.07$ & $ 0.06 \pm 0.06 $\\
$\diff t/t \pm \sigma_{t}$ & $ -0.27 \pm 0.27 $ & $ -0.29 \pm 0.28$ \\
$\diff Y_{0}/Y_{0} \pm \sigma_{Y_{0}}$ & $0.10 \pm  0.14$ & $ 0.16 \pm 0.19 $ \\
$\diff R/R \pm \sigma_{R}$ & $ 0.009 \pm 0.025$ & $ 0.002 \pm 0.023$ \\
\hline
\end{tabular}
\end{table}


\subsection{Differential analysis for HD\,175272 with the full computation of adiabatic frequencies}\label{fullcomputation}

Relations (\ref{dRR}) and (\ref{dMM}), which were used in
Sect.~\ref{Deltanu}, have the advantage of being model
independent. However, scaling relations are known to be
approximate. We accordingly also applied a differential approach
that does not use the scaling relation for $\dnu$. As before, we
differentiated the relations $L=L(M,t,Y_0,Z/X_0)$ and
$\Teff=\Teff(M,t,Y_0,Z/X_0)$ given by the stellar evolutionary
tracks. We now also differentiated the relation
$\dnu=\dnu(M,t,Y_0,Z/X_0)$ given by complete adiabatic
oscillations computations. Finally, we eliminated $\diff L/L$,
which is poorly constrained, by differentiating Stefan-Boltzmann's
law and Eq.~(\ref{numax2}): $\diff L/L = \diff M/M - \diff
\numax/\numax + 7/2\, \diff \Teff/\Teff$. This finally gave the
following linear system (observational constraints are on the
left-hand side and unknowns on the right-hand side):
\begin{eqnarray}
\label{diff3-1}
\frac{\diff \numax}{\numax} -\frac{7}{2}\frac{\diff \Teff}{\Teff} + \frac{\partial \ln L}{\partial \ln Z/X_0}\frac{\diff Z/X_0}{Z/X_0} &=& \nonumber\\
  \left( 1- \frac{\partial \ln L}{\partial \ln M}\right)\frac{\diff M}{M}  - \frac{\partial \ln L}{\partial \ln t}\frac{\diff t}{t} - \frac{\partial \ln L}{\partial \ln Y_0}\frac{\diff Y_0}{Y_0},\\
\frac{\diff \Teff}{\Teff} - \frac{\partial \ln \Teff}{\partial \ln Z/X_0}\frac{\diff Z/X_0}{Z/X_0} &=&\nonumber \\
  \frac{\partial \ln \Teff}{\partial \ln M}\frac{\diff M}{M}
 + \frac{\partial \ln \Teff}{\partial \ln t}\frac{\diff t}{t} + \frac{\partial \ln \Teff}{\partial \ln Y_0}\frac{\diff Y_0}{Y_0}, \\
\label{diff3-3}
\frac{\diff \dnu}{\dnu} -\frac{\partial \ln \dnu}{\partial \ln Z/X_0}\frac{\diff Z/X_0}{Z/X_0}  &=&\nonumber\\
  \frac{\partial \ln \dnu}{\partial \ln M}\frac{\diff M}{M}
  + \frac{\partial \ln \dnu}{\partial \ln t}\frac{\diff t}{t} + \frac{\partial \ln \dnu}{\partial \ln Y_0}\frac{\diff Y_0}{Y_0}.
\end{eqnarray}
The solutions of these equations are given in
Table~\ref{diff2-results}. We explain below how the relative
differences in radius were obtained (Eq.~(\ref{dR})).
Eqs.~(\ref{diff3-1})-(\ref{diff3-3}) can be formulated in matrix
form, which helps in determining the standard errors  on the
parameter relative differences. Let $A_{ij}$ be the $3\times 3$
matrix of the linear system and $x_i=(\diff M/M, \diff t/t, \diff
Y_0/Y_0)$ the three unknowns. With these notations, we have
\begin{equation}
\label{matrixeqn}
x_i = A_{ij}^{-1} b_j = A_{ij}^{-1}B_{jk}\tilde{b}_{k},
\end{equation}
with
\begin{eqnarray}
B_{jk}\tilde{b}_{k} =
\left(
\begin{array}{cccc}
1 & -7/2 & 0 & \displaystyle{\frac{\partial \ln L}{\partial \ln Z/X_0}} \\[0.3cm]
0 & 1 & 0 & -\displaystyle{\frac{\partial \ln T_{\rm eff}}{\partial \ln Z/X_0}} \\[0.3cm]
0 & 0 & 1 & -\displaystyle{\frac{\partial \ln \Delta \nu}{\partial
\ln Z/X_0}}
\end{array}
 \right)
\left(
\begin{array}{c}
\displaystyle{\frac{\diff \numax}{\numax}}  \\[0.3cm]
\displaystyle{\frac{\diff \Teff}{\Teff}}   \\[0.3cm]
\displaystyle{\frac{\diff\dnu}{\dnu}} \\[0.3cm]
\displaystyle{\frac{\diff (Z/X_{0})}{Z/X_{0}}} \end{array}
\right).
\end{eqnarray}
The variances of $x_i$ are obtained assuming independence of the
constraints $\tilde{b}_{k}$, which gives
\begin{equation}
\label{deltax}
\sigma^2_{x_i} = \sum_{j} (A^{-1}_{ij} B_{jk})^2 \sigma^2_{\tilde{b}_{k}}.
\end{equation}
To obtain the relative differences in radii, we simply
differentiated the $\numax$ scaling relation:
\begin{equation}
\label{dR}
\frac{\diff R}{R} = \frac{1}{2}\frac{\diff M}{M} - \frac{1}{4}\frac{\diff \Teff}{\Teff}
- \frac{1}{2}\frac{\diff \numax}{\numax}.
\end{equation}
The term $\diff M/M$ in the above equation is obtained from the
first line of \Eqt~(\ref{matrixeqn}), which gives
\begin{equation}
\label{dRvector} \frac{\diff R}{R} =
\left(\frac{1}{2}(A^{-1}B)_{1} + \left(\frac{-1}{2}, \frac{-1}{4},
0,0\right)\right) \left(
\begin{array}{c}
\displaystyle{\frac{\diff \numax}{\numax}}  \\
\displaystyle{\frac{\diff \Teff}{\Teff}}   \\
\displaystyle{\frac{\diff \dnu}{\dnu}} \\
\displaystyle{\frac{\diff Z/X_{0}}{Z/X_{0}}} \\
\end{array} \right).
\end{equation}
The variance of $\diff R/R$ is thus given by
\begin{equation}
\label{dRvectorsig} \sigma^2_{\diff R / R} =
\left(\frac{1}{2}(A^{-1}B)_{1} + \left(\frac{-1}{2}, \frac{-1}{4},
0,0\right)\right)^2 \left(
\begin{array}{c}
\sigma^2_{\frac{\diff \nu_{\rm max}}{\nu_{\rm max}}}  \\
\sigma^2_{\frac{\diff T_{\rm eff}}{T_{\rm eff}}}   \\
\sigma^2_{ \frac{\diff \Delta \nu}{\Delta \nu}} \\
\sigma^2_{\frac{\diff Z/X_{0}}{Z/X_{0}}} \\ \end{array}
\right).
\end{equation}
Finally, in Table~\ref{175272-direct} and \ref{175272-grid}, our
parameter estimates for the second star with a low SNR are shown,
using the $\dnu$ scaling relation and the full computation of
adiabatic frequencies for the two solar mixtures (GN93 and AGS05),
respectively. Through the differential analysis, it is impossible
to determine the parameter uncertainties of HD\,175272 alone,
because of the correlation between the constraints $y_1$ and
$\delta y = 2(y_2-y_1)/(y_1+y_2)$. Therefore, we proceeded
differently, obtaining the parameter uncertainties by the SVD
method applied to HD\,175272 alone. Using the scaling relation for
$\dnu$, the uncertainties on $R_2$ and $M_2$ are obtained from the
propagation of the uncertainties on the observational constraints,
 \begin{equation}
\frac{\sigma^2_{R_2}}{R_2^2} = \frac{\sigma^2_{\nu_{\rm max,2}}}
{\nu^2_{\rm max,2}} + 4\frac{\sigma^2_{\Delta \nu_{2}}}{\Delta
\nu^2_{2}} +\frac{1}{4} \frac{\sigma^2_{T_{\rm eff,2}}}{T^2_{\rm
eff,2}}
 \end{equation} and
 \begin{equation}
\frac{\sigma^2_{M_2}}{M_2^2} = 9\frac{\sigma^2_{\nu_{\rm max,2}}}
{\nu^2_{\rm max,2}} + 16\frac{\sigma^2_{\Delta \nu_{2}}}{\Delta
\nu^2_{2}} +\frac{9}{4} \frac{\sigma^2_{T_{\rm eff,2}}}{T^2_{\rm
eff,2}}.
 \end{equation}

\begin{table}
\caption{ Parameters of HD\,175272\ obtained by adding the results
of the differential analysis to those obtained for HD\,181420, and
using the $\dnu$ scaling relation.} \label{175272-direct}
\begin{tabular}{lrr}
\hline
\multicolumn{3}{c}{$R/R_{\odot} = 1.63 \pm 0.04$} \\
\multicolumn{3}{c}{$M/M_{\odot} = 1.32 \pm 0.09$} \\
\hline
Solar mixture & GN93 & AGS05 \\
$t_2$(Myr)  & 1627 $\pm$ 251  & 1760 $\pm$ 190 \\
$(Y_{0})_2$ &  0.33 $\pm$ 0.02   & 0.33 $\pm$ 0.02 \\
\hline
\end{tabular}
\end{table}

\begin{table}
\caption{Same as Table~\ref{175272-direct}, but with a full computation
of adiabatic frequencies.} \label{175272-grid}
\begin{tabular}{lrr}
\hline
Solar mixture & GN93 & AGS05 \\
\hline
$R_2/R_{\odot}$  & 1.62      &  1.65 \\
$M_2$      & 1.32 $\pm$  0.36 & 1.29 $\pm$ 0.62 \\
$t_2$(Myr) & 1627 $\pm$ 642  & 1728 $\pm$ 132 \\
$(Y_{0})_2$ & 0.33  $\pm$ 0.16 & 0.34 $\pm$ 0.31 \\
\hline
\end{tabular}
\end{table}

\section{Discussion\label{discussion}}

\subsection{Using $\numax$ in detailed seismic
analysis}\label{nmax}

Using $\numax$ to establish the best model for the reference star,
HD\,181420, was not successful. This failure can have been caused
by possible observational and theoretical uncertainties. Indeed,
$\numax$ may not be the most appropriate quantity for a precise
characterization of stars that are very different from the Sun in
terms of their stellar parameters such as mass, radius,
luminosity, and effective temperature. First of all, from a
theoretical point of view,
 the explanation of the scaling relation $\numax-\nuc$
remains a matter of debate. \cite{2011A&A...530A.142B} have shown
that this scaling relation can be understood by assuming a
resonance at oscillation frequencies around thermal frequency
$1/\tau$ in the super-adiabatic region of the convective envelope.
Nevertheless, in addition to a linear relation between the thermal
frequency and the cutoff frequency $\nuc$, these authors
highlighted a strong dependence with respect to the Mach number
and the parametrization of convection for instance through the
mixing-length parameter. The stars considered in this paper have a
significantly higher effective temperature than the Sun, and
therefore the characteristics of their convective envelope are
very different from that of the Sun: larger Mach numbers and maybe
shorter mixing lengths. Hence, it is not surprising that the
scaling relations calibrated on the Sun yield poor results for
these stars.

Another difficulty that may appear in modeling the reference star
is a poor measurement of $\numax$. The duration of the
observations and thus the limited frequency resolution as well as
the data analysis methods of the stellar oscillation spectrum can
add to the uncertainty on the location of the maximum height in
the oscillation power spectrum envelope
\citep{2011A&A...529A..84B}. The solar-like power spectrum shows
an excess of power in a broad envelope. The profile of this
envelope is assumed to be represented by a Gaussian. Even if this
is the common assumption, it is not based on a firm theoretical
explanation. For example, this shape is clearly not identified for
Procyon \citep[see Figure 10 from][]{2008ApJ...687.1180A}.

However, we were able to use $\numax$ when characterizing a second
star, HD\,175272, with a poor SNR through differential analysis.
Indeed, these two stars, HD\,181420 and HD\,175272, are very
similar and we can assume that this is the same for their
convection zones. Thus, it appears reasonable to accept the
scaling relation $\numax-\nuc$ as a local scaling for stars with
similar convective envelopes.

\subsection{Choosing a reference}

HD\,175272, with an EACF just above the threshold level for
detecting solar-like oscillations, corresponds typically to a
low-quality oscillation spectrum \citep[Table 3
of][]{2009A&A...508..877M}. As a comparison, the reference
HD\,181420 has an EACF of about 240, high enough to allow a
precise identification of the modes over more than nine radial
orders. Even though \cite{2009A&A...506...51B} proposed two
options for the mode identification because this star may be
affected by the HD\,49933 misidentification syndrome
\citep{2008A&A...488..705A,2009A&A...507L..13B}, the ridge
identification in HD\,181420 is unambiguous
\citep{2009A&A...508..877M,2012ApJ...751L..36W}. Much higher EACF
are easily observed in CoRoT and \Kepler\ targets, such as
HD\,49933 \citep{2009A&A...507L..13B}, HD\,49385
\citep{2010A&A...515A..87D}, or the solar analogs 16 Cyg A and B
\citep{2012ApJ...748L..10M}. This means that there are a large
number of main-sequence stars and subgiants that can serve as
seismic references. We consider that an EACF of 100 is enough for
the reference, so that we currently identify more than 80 possible
references for subgiants and main-sequence stars, according to
previous CoRoT and \Kepler\ observations
\citep[e.g.,][]{2011MNRAS.415.3539V}. The only domain where the
set of reference stars appears to be loose in the main sequence
for stars with a lower mass than the Sun is for $\numax \ge
3.6$\,mHz. In the red giant domain, the number of potential
references is huge because it benefits from long observation runs
with both CoRoT and \Kepler\
\citep[e.g.,][]{2010A&A...517A..22M,2011MNRAS.415.3783D,2013ApJ...765L..41S}.

\section{Conclusion\label{conclusion}}

We have presented a new approach to determine stellar properties
of solar-like stars that we call differential seismology of twins.
This method makes it possible to constrain the global
characteristics of stars with a low SNR from reference stars
observed with a higher SNR.

We applied this method to two CoRoT solar-like stars: HD\,181420
with a high SNR oscillation spectrum served as a reference for
modeling the secondary target of the first CoRoT short run,
HD\,175272. This opened a positive perspective for the analysis of
low SNR asteroseismic data from the CoRoT and Kepler missions, or
for those observed during a short period of time, such as the
Kepler one-month time series \citep{2011Sci...332..213C}. These
targets can benefit from a comparative analysis. This can be the
case for targets showing a peculiar interest, such as members of a
double system or stars hosting an exoplanet.

To obtain information on the less-known star HD\,175272 from the
well-known reference star HD\,18142, we first found the best
stellar  model of HD\,181420. We note that a difficulty appeared
when we tried to find the best model of the reference star taking
into account all observational constraints including $\numax$. In
fact, the two stars considered in this study, HD\,181420 and
HD\,175272, are very different from the Sun in terms of their
mass, radius, luminosity, and effective temperature and therefore
their seismic properties. This difficulty might originate from
either a departure from the linear relation between $\numax$ and
$\nuc$ or an inaccurate measurement of $\numax$.

Then, we performed a differential analysis to characterize the
lower SNR target HD\,175272 based on its seismic similarity with
the reference star HD\,181420. Although the calibration relying on
$\numax$ is not appropriate for the reference star HD\,181420
compared with the Sun, the two stars HD\,181420 and HD\,175272 are
so close to each other that the scaling relation can be used
locally for nearby stars. We therefore decided to assume here that
the physical mechanism is the dominant factor in the failure to
scale $\numax$ from the Sun to HD\,181420.

The results of our differential analysis presented in Tables
\ref{diff1-results} and \ref{diff2-results} show that the standard
errors are significant compared with the relative differences.
This simply results from the large measurement errors given in
Table~\ref{diffobs}. The differential approach decreases the
inaccuracies of the forward seismic analysis, but the precision of
the results remains intrinsically related to the precision of the
measurements. The very large standard error found for $\diff
Y_0/Y_0$ is striking. We checked carefully that this is indeed the
correct result of our analysis. Degeneracies are often present in
the stellar parameters -- constraints relation, so that some
parameters cannot be determined precisely. This is the case for
$Y_0$ in the specific region of the Hertzsprung-Russel diagram
corresponding to the twin stars.

Comparing the results found using the scaling relation for $\dnu$
(Table~\ref{diff1-results}) with those obtained from a full
computation of adiabatic frequencies (Table~\ref{diff2-results})
shows that they are very close. This clearly shows that the
inaccuracy of the $\dnu$ scaling relation does not affect the
results of the differential analysis significantly. For
measurement precisions similar to those of this study, it is
therefore fully justified to use Eqs. (\ref{dRR}) and (\ref{dMM})
for an easy and rapid determination of the radius and mass
relative differences. Comparing the results obtained with the two
solar mixtures gives a lower bound on the inaccuracies of the
differential seismic study. Here, the inaccuracies resulting from
the choice of the solar mixture are smaller than the imprecisions
resulting from the measurements, but not negligible.

With our differential method, the scientific output of many
asteroseismic objects with a poor SNR might benefit from the
accurate modeling of nearby reference stars with a high SNR. Owing
to the large number of asteroseismic targets observed with a high
SNR in the many regions of the Hertzsprung-Russel diagram where
solar-like oscillations are present, this type of differential
seismic analysis allows us to characterize a large number of
different types of stars  with a low SNR, from red giants to
main-sequence stars, and to enhance the precision of the
asteroseismic output. This method is not just useful for
characterizing the lower SNR targets. It can can also be applied
to very well constrained stars. In this case, it would give a very
precise determination of the structural differences between nearby
stars. The strength of the differential method is here that the
results are less sensitive to the systematic errors coming from
both the modeling and the data analysis method.

\begin{acknowledgements}
NO acknowledges support from the Scientific and Technological
Research Council of Turkey (TUBITAK). RAG acknowledges the
financial support of the CNES/CoRoT grant at the SAp.
\end{acknowledgements}

\listofobjects

\bibliographystyle{aa} 
\bibliography{biblio175272} 

\end{document}